\documentclass[12pt]{article}
\usepackage[english]{babel}
\usepackage[utf8x]{inputenc}

\usepackage{latexsym}
\usepackage{amsfonts}
\usepackage{amsmath}
\usepackage{amssymb}
\usepackage{footnote}

\usepackage{graphicx}
\usepackage[colorinlistoftodos]{todonotes}
\usepackage{dsfont}
\usepackage{ mathrsfs }
\usepackage[colorlinks=true, allcolors=blue]{hyperref}
\DeclareGraphicsExtensions{.pdf, .png, .jpg}
\newcommand{\E}{\mathrm{e}}
\newcommand{\I}{\mathrm{i}}

\newcommand{\be}{\begin{equation}}
\newcommand{\ee}{\end{equation}}
\newcommand{\bean}{\begin{eqnarray*}}
	\newcommand{\eean}{\end{eqnarray*}}

\newcommand{\bea}{\begin{eqnarray}}
\newcommand{\eea}{\end{eqnarray}}

\usepackage{bbding} 

\usepackage{indentfirst}

\usetikzlibrary{shapes.misc}

\tikzset{cross/.style={cross out, draw=black, minimum size=2*(#1-\pgflinewidth), inner sep=0pt, outer sep=0pt},
cross/.default={1pt}}

\title{Digital representation of continuous observables in Quantum Mechanics}
\author{M. G. Ivanov, A. Yu. Polushkin\\
\small Moscow Institute of Physics and Technology}

\begin{document}
\maketitle
\abstract{
To simulate the quantum systems at classical or quantum computers, it is necessary to reduce continuous observables (e.g. coordinate
and momentum or energy and time) to discrete ones. In this work we consider
the continuous observables represented in the positional systems as a series of powers of the radix mulitplied over the
summands (``digits``), which turn out to be Hermitean operators with discrete
spectrum. We investigate the obtained quantum mechanical operators of
digits, the commutation relations between them and the effects of choice of
numeral system on the lattices and representations. Furthermore, during the
construction of the digital representation renormalizations of diverging sums
naturally occur.}

\section{Introduction}

Quantum computations is a very perspective and dynamically developing area of science. 
Great hopes are placed on quantum computers, primarily due to the expected ability to solve problems that are fundamentally impossible to compute on classical computers due to the asymptotic complexity.
Despite the fact, that the main stress is usually placed  on the problems of quantum cryptography \cite{qc_persp}, it is obvious that since the quantum key distribution and post-quantum cryptography is safe \cite{shohr_preskill}, the main field of applications for the quantum computer tends to be the ``peaceful`` modelling of complex systems, according to the original Feynman's idea \cite{Feynman}. Moreover, the current development of the technologies in the field of quantum computers already enables to solve some problems, which are problematic to be solved on the classical computer. The fields of application are also  wide -- from chemistry \cite{quant_chem} to ``industrial`` problems \cite{quant_industrial}.

In computational problems, it becomes necessary to discretize continuous observables. 
The representation considered in the paper can be useful for modeling both classical and quantum systems on both classical and quantum computers. 

In this work the representation of continuous quantum observables as a row of powers of observables-digits is considered. 
We consider the expansions of the coordinate and momentum into a series in terms of the powers of the base of the corresponding numeral systems, and the digit itself also turns out to be observable. 
The momentum operator is introduced as a generator of coordinate shifts. As a result the Fourier expansion in this formalism is equivalent to a series expansion in terms of shift operators.
Explicit expressions are obtained for all systems. Moreover, in the work are considered commutation relations, which also turn out not to be trivial. 

Among other things, when constructing this representation, a number of interesting ``physical`` effects naturally arise. 
For example, we demonstrate the connection between the choice of a binary digit, gauge transformations and the Aharonov–Bohm effect.

As a part of the research work, various renormalization procedures also naturally arose, allowing one to assign finite values to some divergent series, which may be useful in the context of quantum field theory.

This paper presents the generalization of the digital representation \cite{bin-ivanov} and \cite{ter-bin}, for $q = 2$ and $q = 3$ to arbitrary $q \geq 2$. To make it easier to compare the results of previously published articles to the newest one, the style and contest retained where possible, which sometimes results in blocks of self-citation. 
Despite the fact that the generalization from $q \in \{2,3\}$ to the integer $q = n$ was quite straightforward, the particulary interesting are the ``shifted`` lattices with non-integer digits, which were not covered in previous papers properly, but induce interesting effects, such as non-trivial boundary conditions, providing the analogy of the Aharonov–Bohm effect.

\section{Lattice definition}

To make the notation more compact, we consider here and hereafter the system of units, where the Planck constant is equal to one. In other words, we assume $h = 1$ and hence $1/\hbar = 2\pi$. While speaking about quantum observables we are going to concentrate primary on coordinate and momentum, following the notation of the original articles \cite{bin-ivanov}, \cite{ter-bin}, but it is clear, that we can think about them as about any pair of Fourier-conjugated quantum observables, e.g. time and energy.

\subsection{Coordinate lattice}

We assume that the coordinate is described by $q$ digit-qudit and the coordinate lattice consists of $N = q^n$ nodes, which we assume to be cyclic (after the last one comes the first). If we suppose that the \textit{coordinate lattice constant} is $\Delta x = q^{-n_{-}}$, then the \textit{lattice period} is equal to $\Xi = N\Delta x = q^{n_{+}}$, $n_{+} = n - n_{-}$. We assume that the values $x$ range from 0 to $\Delta x \cdot (N - 1)$.

On the coordinate lattice, a natural addition operation is induced from $\mathds{Z}_{N}$, for which $x = x + \Xi$. It is possible to use other representations of the lattice $\Delta x \cdot \mathds{Z}_{N}$ by real numbers. For example, taking the equivalence of $x$ and $x + \Xi$ into account, we further need a representation in which $x$ ranges from $-\Delta x \cdot (N-1)/2$ to $+\Delta x \cdot (N-1)/2$. The power series for the coordinate on a finite lattice is finite:
\begin{equation}
    x = \sum\limits_{s = -n_{-}}^{n_{+} - 1}x_{s}q^{s} = \sum\limits_{s = -n_{-}}^{n_{+} - 1}\mathbf{d}(s,x)q^{s}.
\end{equation}
Here, $x_{s} = \mathbf{d}(s,x)$ is the $s$-th digit in the {\bf d}igital expansion of $x$. 

Any function of the observable can be properly defined in the representation, in which the operator of the observable is diagonal. For instance, the operator $\theta(\hat{x})$ is frequently used. This enables us to define $\hat{x}_s$ in coordinate representation and $\hat{p}_r$ in momentum representation without applying any additional procedures.

We sometimes specify a range of powers of three that defines a lattice and write $x_{s} = \mathbf{d}_{n_{-}n_{+}}(s,x)$.
 
We introduce the coordinate basis $\{|x\rangle \}_{x \in \Delta x \cdot \mathds{Z}_{N}}$ for functions defined on the lattice:
\begin{equation}
    \hat{x}|x\rangle = x|x\rangle, \quad \langle x^{\prime}|x^{\prime \prime} \rangle = \delta_{x^{\prime},x^{\prime \prime}}, \quad \psi(x) = \langle x|\psi \rangle, \quad x \in \Delta x \cdot \mathds{Z}_{N}.
\end{equation}
We represent wave functions (ket vectors) in the forms of columns whose rows are ordered in decreasing order of $x$. Thus, if $x$ varies from $0$ to $(N - 1)\Delta x$, then
\begin{equation}\psi (x) =\left(
    \begin{array}{c}
           \psi((N - 1)\Delta x)\\
           \psi((N - 2)\Delta x)\\
           .\\
           .\\
           .\\
            \psi(\Delta x)\\
            \psi(0)\\
    \end{array}\right).
\end{equation}

\subsection{Momentum lattice.}
We define the momentum operator $\hat{p}$ as the generator of the cyclic shifts $\widehat{T}_{A}$ along the coordinate lattice:
\begin{equation}
    \widehat{T}_{A} \psi(x) = \psi(x + A), \qquad \widehat{T}_{A} =\E^{2\pi \I A \hat{p}}, \quad A \in \Delta x \cdot \mathds{Z}. 
\end{equation}
Such operators were considered on Weyl's classic book \cite{Weyl} and more detailed later by Schwinger \cite{Schwinger}. 
Here we have to mention that as we have the standard operators $\hat{x}$ and $\hat{p}$, which are generators of momentum and coordinate shifts respectively: $\E^{2\pi i A \hat{p}} = \widehat{T}_{A}, \E^{2 \pi i B \hat{x}} = \widehat{S}_{B}$, the operators of digits are the functions of the corresponding observables: $\hat{x}_s = f(\hat{x},s), \hat{p}_r = g(\hat{p},r)$. This leads to the fact that the algebra of the observables set by $\hat{x}_{s}$ and $\hat{p}_{r}$ is equivalent to the one set by coordinate and momentum. In this paper we are not considering quantum deformed algebras (see, for instance, \cite{q_def_1}, \cite{q_def_2}), which could be the object of further research.

Because the coordinate lattice is periodic, the shift by the period $\Xi$ must be identity transformation, i.e., for eigenvalues of operator $\hat{p}$, we have 
$\Xi\cdot p\in\mathds{Z}$.
This gives the \textit{momentum step} $\Delta p$,
\begin{equation}
    \Xi \cdot \Delta p = 1, \quad \Delta p = q^{-n_{+}}, \quad \Delta p \cdot \Delta x = \frac{1}{N} = q^{-n}.
\end{equation}
The number of points in the spectrum of momentum is the same as for coordinate,
i.e. for momentum,
we have a periodic lattice with the same number of nodes but a different period $\Pi = \Delta p \cdot N = q^{n_{-}}$, $\Pi \Xi = N$. 
The momentum lattice is denoted by $\Delta p \cdot \mathds{Z}_{N}$. 
The power series for the momentum is also finite:
\begin{equation}
    p = \sum\limits_{r = -n_{+}}^{n_{-} -1}p_{r}q^{r} = \sum\limits_{r = -n_{+}}^{n_{-} -1} \mathbf{d}(r,p)q^r.
\end{equation}
Here, $p_r=\mathbf{d}(r,p)$ is the $r$-th digit in the \textbf{d}igital expansion of $p$.
We sometimes specify a range of powers of $q$ that defines a lattice and write $p_r = \mathbf{d}_{n_+n_-}(r,p)$.

\subsection{Minimum shift.}
The minimum shift $\widehat{T}_{\Delta x}$ is a shift by the lattice step $\Delta x$; any other shift on a given lattice is a power $\widehat{T}_{A} = (\widehat{T}_{\Delta x})^{A/\Delta x}$, where $A/\Delta x \in \mathds{Z}_{N}$:
\begin{equation}\label{T1}
    \widehat{T}_{A}\psi(x) = \psi(x + A), \, \widehat{T}_A|x\rangle = |x - A\rangle, \, 
    \langle x^{\prime}|\widehat{T}_{A}|x^{\prime \prime}\rangle = \delta_{x^{\prime},x^{\prime \prime }- A} = \delta_{x^{\prime} + A,x^{\prime \prime}}.
\end{equation}
Moreover,
\begin{equation}\label{T2}
    \hat T_{\Delta x}\psi(x)=\hat T_{\Delta x}
\left(\begin{array}{c}
\psi((N-1)\Delta x)\\
\psi((N-2)\Delta x)\\
\vdots\\
\psi(2\Delta x)\\
\psi(\Delta x)\\
\psi(0)
\end{array}\right)=
\left(\begin{array}{c}
\psi(0)\\
\psi((N-1)\Delta x)\\
\psi((N-2)\Delta x)\\
\vdots\\
\psi(2\Delta x)\\
\psi(\Delta x)
\end{array}\right)=\psi(x+\Delta x).
\end{equation}
The sum $x + \Delta x$ is taken in the sense $x \in \Delta x \cdot \mathds{Z}_{N}$, i.e., this is a cyclic shift of the function on the lattice down one position.
 
The eigenvalues of the minimum shift operator are $N$th roots of unity and are related to the eigenvalues of the momentum operator (which has not yet been introduced explicitly):
\begin{equation}
    \lambda^{N} = 1, \qquad \lambda_{p} = \E^{2\pi \I \Delta x p} =  \E^{2\pi \I \Delta x \Delta p p/\Delta p} = (\lambda_{\Delta p})^{p/\Delta p},
\end{equation}
where we take $\Delta x \Delta p = 1/N$ and $p/\Delta p \in \mathds{Z}_{N}$ into account and
\begin{equation}
    \Lambda = \lambda_{\Delta p} = \E^{2\pi \I \Delta p \Delta x} = \E^{2\pi \I /N}.
\end{equation}
The corresponding eigenvalues are obtained from the relation $\psi(x) = \widehat{T}_x\psi(0)$. The normalized eigenvectors have the forms
\bea
    \psi_{\lambda_{p}}(x^{\prime}) &=& \langle x^{\prime}|\psi_{\lambda_{p}}\rangle = \frac{\lambda_{p}^{x^{\prime}/\Delta x}}{\sqrt{N}} = \frac{\E^{2\pi \I x^{\prime} p}}{\sqrt{N}},\\ 
    \nonumber
    \langle \psi_{\lambda_{p}}|x^{\prime \prime}\rangle &=& \langle x^{\prime \prime}|\psi_{\lambda_{p}}\rangle^{*} = \frac{\lambda_{p}^{-x^{\prime \prime}/\Delta x}}{\sqrt{N}} = \frac{\E^{-2\pi \I x^{\prime \prime}}}{\sqrt{N}}.
\eea
We can write the projector on the (one-dimensional) eigensubspace of the operator $\widehat{T}_{\Delta x}$ as
\bea
    \widehat{P}_{\lambda_{p}} &=& |\psi_{\lambda_{p}}\rangle\langle\psi_{\lambda_{p}}|,\\ \nonumber
    \langle x^{\prime}|\widehat{P}_{\lambda_{p}}|x^{\prime \prime} \rangle &=& \frac{\lambda_{p}^{(x^{\prime} - x^{\prime \prime})/\Delta x}}{N} = \frac{\lambda_{p}^{d/\Delta x}}{N} = \frac{\E^{2\pi \I p d}}{N}, \quad d = x^{\prime} - x^{\prime \prime}.
\eea
In this notation $x^{\prime}$ labels rows of matrix, and $x^{\prime\prime}$ labels columns.
 
The eigenstates of the minimum shift operator are also eigenstates of the momentum operator and can be hence written differently:
\begin{equation}
    |\psi_{\lambda_{p}}\rangle = |\psi_{p}\rangle = |p\rangle, \quad \langle x | p \rangle = \frac{\E^{2\pi \I x p}}{\sqrt{N}}.
\end{equation}

\subsection{Group of shifts.} 

We make a trivial remark that might nevertheless be of some interest for an arbitrary positional number system. 
We constructed the momentum operator such that it generates a symmetry group with respect to the shifts of the coordinate lattice by an integer number of nodes, i.e. a group isomorphic to the group (with respect to addition) of the residues modulo division by $N:\, \Delta x \cdot \mathds{Z}_{N} \approx \mathds{Z}_{N}$. But we can consider unitary operators of the form $\widehat{T}_{A} = e^{2\pi i A \hat{p}}, \, A \in \mathds{R}.$ Such operators correspond to the cyclic shifts by an arbitrary value (not necessarily a multiple of $\Delta x$). The corresponding group is isomorphic to the group $\mathds{R} / \Xi \approx SO(1) \approx U(1)$ of rotations of a circle by an arbitrary angle. Addition is again understood modulo $\Xi$, ($A = A + \Xi$). In the case $\Xi = \infty$, the group of symmetries coincides with the group $\mathds{R}$ of real numbers with respect to addition.
 
We see that if the Hamiltonian on the lattice is expressed in terms of the operator $\hat{p}$, then the presence of the lattice does not violate translation invariance under arbitrary translations (not necessarily by an integer number of lattice sites), but the operator $\hat{p}$ (as we see below) turns out to be nonlocal, i.e., matrix elements $\langle x^{\prime} | \hat{p} | x^{\prime\prime} \rangle$ can be nonzero for arbitrary large values $x^{\prime} - x^{\prime \prime}$ (in the lattice).
 
We can specify a state $|x_{0}\rangle = \widehat{T}_{-x_{0}}|0\rangle$ with an arbitrary value $x_{0} \notin \Delta x \cdot \mathds{Z}_{N}$, but such a state is not a state with a certain value of the coordinate, because it decomposes into several basic states $\{ |x\rangle \}_{x \in \Delta x \cdot \mathds{Z}_{N}}$.


It is also noteworthy, that in case of $integer$ shifts (such that $A = \Delta x \cdot n, B = \Delta p \cdot m; \,  n,m \in \mathds{N}$, the Weyl's canonical relation  for the group commutator is observed: 
\be
\widehat{T}_{a}\widehat{S}_{b}\widehat{T}_{-a}\widehat{S}_{-b} = \E^{2 \pi \I a b},
\ee
while for ``non-integer`` shifts this relation is violated. Indeed, for arbitrary $b \neq \Delta p \cdot \mathbb{Z}_N$ we can express the ``non-integer`` shift through the ``integer`` ones:
\be
  \widehat S_b=\frac{1-\E^{2\pi\I\Delta x\cdot  bN}}{N}\sum_{B\in\Delta p\cdot\mathbb{Z}_N}
  \frac{\E^{-2\pi\I x_0 (B-b) }\widehat S_B}{1-\E^{2\pi\I\Delta x\cdot (b-B)}}.
\ee
Here $x_0$ is the value of shift of the coordinate lattice (in other words, ``the least number`` on the lattice). Hence, for the group commutator we obtain the following expression:
\be
  \widehat T_a\widehat S_b\widehat T_{-a}\widehat S_{-b}=\left(\frac{2\sin(\pi\Delta x\cdot  b N)}{N}\right)^2
  \sum_{B_1,B_2\in\Delta p\cdot\mathbb{Z}_N}
  \frac{\E^{-2\pi\I x_0(B_1-B_2)}
  	\E^{2\pi\I a B_1}\,\widehat S_{B_1-B_2}}{(1-\E^{2\pi\I\Delta x\cdot (b-B_1)})(1-\E^{-2\pi\I\Delta x\cdot (b-B_2)})}.
\ee
Analogically in the case of $a \neq \Delta x \cdot \mathbb{Z}_N$ for the coordinate shift we can write:
\be
  \widehat T_a=\frac{1-\E^{2\pi\I\Delta p\cdot  aN}}{N}\sum_{A\in\Delta x\cdot\mathbb{Z}_N}
  \frac{\E^{-2\pi\I p_0 (A-a) }\widehat T_A}{1-\E^{2\pi\I\Delta p\cdot (a-A)}}.
\ee
Here $p_0$ is the value of shift of the momentum lattice (in other words, ``the least number`` on the lattice). And the group commutator has the form:
\be
  \widehat T_a\widehat S_b\widehat T_{-a}\widehat S_{-b}=\left(\frac{2\sin(\pi\Delta p\cdot  a N)}{N}\right)^2
  \sum_{A_1,A_2\in\Delta x\cdot\mathbb{Z}_N}
  \frac{\E^{-2\pi\I p_0(A_1-A_2)}
  	\E^{ 2 \pi\I b A_2}\,\widehat T_{A_1-A_2}}{(1-\E^{2\pi\I\Delta p\cdot (a-A_1)})(1-\E^{-2\pi\I\Delta p\cdot (a-A_2)})}.
\ee
\section{Operators of digits and their decomposition by shifts}

We defined the momentum operator such that the Fourier harmonic of the momentum is given by the operator $\widehat{T}_{A} = \E^{2\pi \I A \hat{p}}$ of the coordinate shift. Therefore, if we take Fourier transform for the momentum digits
\begin{equation}
    \mathbf{d}_{n_{+}n_{-}}(r,\hat{p}) = \sum\limits_{A \in \Delta x \cdot \mathds{Z}_{N}}\tilde{\mathbf{d}}_{n_{+}n_{-}}(r, A)\E^{2\pi \I A \hat{p}},
\end{equation}
then we obtain the decomposition of the momentum digit by coordinate shifts%
\begin{equation}
   \mathbf{d}_{n_{+}n_{-}}(r, \hat{p}) = \sum\limits_{A \in \Delta x \cdot \mathds{Z}_{N}} \tilde{\mathbf{d}}_{n_{+}n_{-}}(r, A) \widehat{T}_{A}.
\end{equation}

\subsection{Classical positional numeral systems}

\subsubsection{Base-q numeral system}

Let's consider the numeral system with digits $\{0,1, \hdots, q\}$. After computations we obtain the following expression for decomposition of the momentum digit over coordinate shifts:
\be
\hat{p}_{r} = \frac{q - 1}{2} \hat{1} - \Delta p q^{-r} \sum\limits_{D \in \mathbb{Z}(q^r/\Delta p)} \sum\limits_{\sigma = 1}^{q - 1}\frac{\widehat{T}_{-A}}{1 - \exp(2\pi \I \Delta p A)}, \, A = q^{-r}(D + \sigma/q).
\ee
For decomposition of the coordinate in this case we have:
\be
\hat{x}_{s} = \frac{q - 1}{2} \hat{1} - \Delta x q^{-s}   \sum\limits_{D \in \mathbb{Z}(q^s/\Delta x)} \sum\limits_{\sigma = 1}^{q - 1}\frac{\widehat{S}_{B}}{1 - \exp(2\pi \I \Delta x B)}, \, B = q^{-s}(D + \sigma/q).
\ee
Now let's turn to the examples, considered in previous papers.

\subsubsection{Binary non-symmetric system}

The first system to be considered is the ``classical`` binary system, which was being discussed in the original paper \cite{bin-ivanov}. This system has digits $\{0,1\}$ and will be called \textit{``binary non-symmetric system``}.

\begin{figure}[h]
\label{bns_lattice}
\centering
\begin{tikzpicture}
\draw [->] (-5,0) -- (5,0) node [below] {$x$};
\draw [-|] (0,0) -- (1,0) node [below] {1};
\draw [-|] (0,0) -- (0,1) node [right] {1};
\draw [->] (0,-2) -- (0,2) node [right] {$\mathbf{b_{ns}}(s,x)$};

\fill[green] (-4,0) circle (2pt);
\fill[green] (-3.5,0) circle (2pt);

\fill[green] (-3,1) circle (2pt);
\fill[green] (-2.5,1) circle (2pt);

\fill[green] (-2,0) circle (2pt);
\fill[green] (-1.5,0) circle (2pt);

\fill[green] (-1,1) circle (2pt);
\fill[green] (-0.5,1) circle (2pt);

\fill[green] (0,0) circle (2pt);
\fill[green] (0.5,0) circle (2pt);

\fill[green] (1,1) circle (2pt);
\fill[green] (1.5,1) circle (2pt);

\fill[green] (2,0) circle (2pt);
\fill[green] (2.5,0) circle (2pt);

\fill[green] (3,1) circle (2pt);
\fill[green] (3.5,1) circle (2pt);

\draw[blue] (0,0) circle (2pt);
\draw[blue] (0.5,0) circle (2pt);
\draw[blue] (1,0) circle (2pt);
\draw[blue] (1.5,0) circle (2pt);

\draw[blue] (2,1) circle (2pt);
\draw[blue] (2.5,1) circle (2pt);
\draw[blue] (3,1) circle (2pt);
\draw[blue] (3.5,1) circle (2pt);

\draw[blue] (-0.5,1) circle (2pt);
\draw[blue] (-1,1) circle (2pt);
\draw[blue] (-1.5,1) circle (2pt);
\draw[blue] (-2,1) circle (2pt);

\draw[blue] (-2.5,0) circle (2pt);
\draw[blue] (-3,0) circle (2pt);
\draw[blue] (-3.5,0) circle (2pt);
\draw[blue] (-4,0) circle (2pt);

\end{tikzpicture}
\caption{Plot of the value of the binary digit number $s$ on a lattice for an ``non-symmetric system'', ($n_{-} = 1$), $s = 0$ -- green filled circles, $s = 1$ -- blue circles.}
\end{figure}
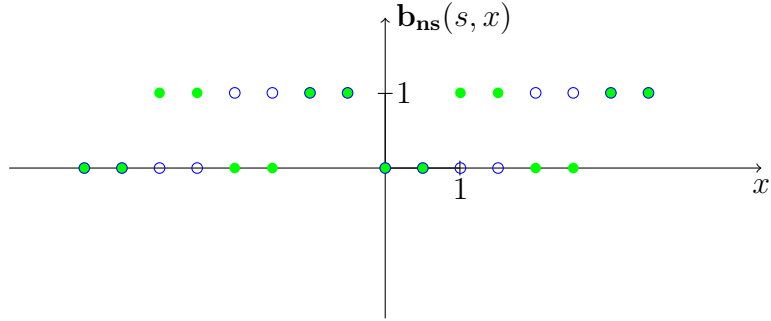

In this case we obtain the following result:

\begin{equation}
\hat{p}_r = \frac{1}{2} \hat{1} - \Delta p \, 2^{-r} \; \sum\limits_{D \in \mathds{Z}_{2^{r}/\Delta p}} \frac{\widehat{T}_{-A}}{1 - \exp(2\pi \I \Delta p A)}, \, A=2^{-r}(D + 1/2).
\end{equation}

\subsubsection{Ternary non-symmetric system}
In paper \cite{ter-bin} was introduced the ternary system with digits $\{0,1,2\}$, called \textit{``ternary non-symmetric system``}.

\begin{figure}[h!]
    \centering
\begin{tikzpicture}
\draw [->] (-6,0) -- (6,0) node [below] {$x$};
\draw [-|] (0,0) -- (1,0) node [below] {1};
\draw [-|] (0,0) -- (0,1) node [right] {1};
\draw [->] (0,-0.5) -- (0,3) node [right] {$\mathbf{t_{ns}}(s,x)$};

\fill[blue] (-6,0) circle (2pt);
\fill[blue] (-5.666,0) circle (2pt);
\fill[blue] (-5.333,0) circle (2pt);

\fill[blue] (-5,1) circle (2pt);
\fill[blue] (-4.666,1) circle (2pt);
\fill[blue] (-4.333,1) circle (2pt);

\fill[blue] (-4,2) circle (2pt);
\fill[blue] (-3.666,2) circle (2pt);
\fill[blue] (-3.333,2) circle (2pt);

\fill[blue] (-3,0) circle (2pt);
\fill[blue] (-2.666,0) circle (2pt);
\fill[blue] (-2.333,0) circle (2pt);

\fill[blue] (-2,1) circle (2pt);
\fill[blue] (-1.666,1) circle (2pt);
\fill[blue] (-1.333,1) circle (2pt);

\fill[blue] (-1,2) circle (2pt);
\fill[blue] (-0.666,2) circle (2pt);
\fill[blue] (-0.333,2) circle (2pt);

\fill[blue] (0,0) circle (2pt);
\fill[blue] (0.333,0) circle (2pt);
\fill[blue] (0.666,0) circle (2pt);

\fill[blue] (1,1) circle (2pt);
\fill[blue] (1.333,1) circle (2pt);
\fill[blue] (1.666,1) circle (2pt);

\fill[blue] (2,2) circle (2pt);
\fill[blue] (2.333,2) circle (2pt);
\fill[blue] (2.666,2) circle (2pt);

\fill[blue] (3,0) circle (2pt);
\fill[blue] (3.333,0) circle (2pt);
\fill[blue] (3.666,0) circle (2pt);

\fill[blue] (4,1) circle (2pt);
\fill[blue] (4.333,1) circle (2pt);
\fill[blue] (4.666,1) circle (2pt);

\fill[blue] (5,2) circle (2pt);
\fill[blue] (5.333,2) circle (2pt);
\fill[blue] (5.666,2) circle (2pt);

\draw[red] (0,0) circle (2pt);
\draw[red] (0.333,0) circle (2pt);
\draw[red] (0.666,0) circle (2pt);
\draw[red] (1,0) circle (2pt);
\draw[red] (1.333,0) circle (2pt);
\draw[red] (1.666,0) circle (2pt);
\draw[red] (2,0) circle (2pt);
\draw[red] (2.333,0) circle (2pt);
\draw[red] (2.666,0) circle (2pt);

\draw[red] (3,1) circle (2pt);
\draw[red] (3.333,1) circle (2pt);
\draw[red] (3.666,1) circle (2pt);
\draw[red] (4,1) circle (2pt);
\draw[red] (4.333,1) circle (2pt);
\draw[red] (4.666,1) circle (2pt);
\draw[red] (5,1) circle (2pt);
\draw[red] (5.333,1) circle (2pt);
\draw[red] (5.666,1) circle (2pt);

\draw[red] (-0.333,2) circle (2pt);
\draw[red] (-0.666,2) circle (2pt);
\draw[red] (-1,2) circle (2pt);
\draw[red] (-1.333,2) circle (2pt);
\draw[red] (-1.666,2) circle (2pt);
\draw[red] (-2,2) circle (2pt);
\draw[red] (-2.333,2) circle (2pt);
\draw[red] (-2.666,2) circle (2pt);
\draw[red] (-3,2) circle (2pt);

\draw[red] (-6,1) circle (2pt);
\draw[red] (-3.333,1) circle (2pt);
\draw[red] (-3.666,1) circle (2pt);
\draw[red] (-4,1) circle (2pt);
\draw[red] (-4.333,1) circle (2pt);
\draw[red] (-4.666,1) circle (2pt);
\draw[red] (-5,1) circle (2pt);
\draw[red] (-5.333,1) circle (2pt);
\draw[red] (-5.666,1) circle (2pt);

\end{tikzpicture}
    \caption{A plot of the value of the ternary digit number $s$ for a finite lattice ($n_{-} = 1$), $s=0$ -- blue filled circles, $s=1$ -- red circles.}
\end{figure}
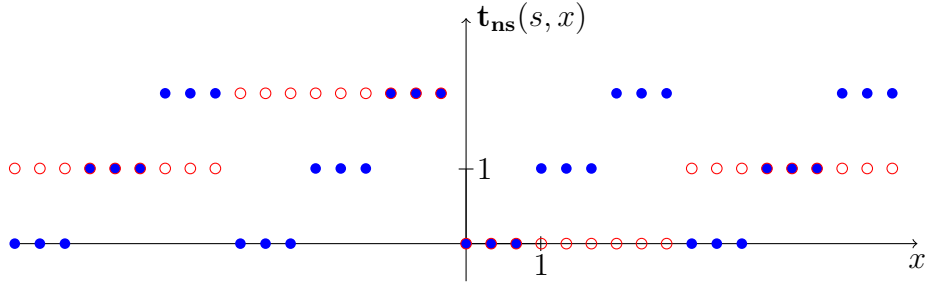
For this system we easily get the following expression:
\begin{equation}
    \hat{p}_r =  \Delta p\, 3^{-r}\sum\limits_{D \in \mathds{Z}_{3^r/\Delta p}} \sum\limits_{\sigma = 1}^{2}  \frac{(-1)^{D + \sigma}}{2 \I \sin\left(\pi\,\Delta p \,A\right)}\widehat{T}_{-A}, \, 
    A=3^{-r}(D + \sigma/3).
\end{equation}
\subsection{Shifted positional systems}

\subsubsection{Base-q numerical system on the shifted lattice}
Let's now consider the positional system with digits $\{d_1, d_1 + 1, \hdots, d_1 + q - 1\}$, here $d_1$ may be non-integer. The shift of the least digit induces the shift of the lattice. Indeed, the ``least`` number on such lattice $(d_1d_1\hdots d_1, d_1 \hdots d_1)$ can be represented as following expression:
\begin{equation} \label{renorm}
	\sum\limits_{s = -n_{-}}^{n_{+} - 1 } q^{s} d_{1} \,  (\text{mod} \, q^{n_{+}})= -q^{-n_{-}} d_{1} = -d_{1} \Delta x,
\end{equation}
which means that we obtain the shifted on $-\Delta x d_1$ lattice and in general case zero is not obliged to appear the node of the lattice.
For this case we can write the general expression for the momentum digit:
\be
\hat{p}_r = \frac{d_1 + q - 1}{2} \; \hat{1} - \Delta p \, q^{-r} \sum\limits_{D \in \mathds{Z}_{q^r/\Delta p} } \sum\limits_{\sigma = 0}^{q - 1} \frac{\widehat{T}_A }{1 - \exp (2 \pi \I \Delta p  \, A)}  \exp(- 2 \pi \I  \Delta p \, A d_1),
\ee
$\, A = q^{-r}(D + \sigma/q),$ and the following expression for the coordinate digit:
\be
\hat{x}_s = \frac{d_1 + q - 1}{2} \; \hat{1} - \Delta x \, q^{-s} \sum\limits_{D \in \mathds{Z}_{q^s/\Delta x} } \sum\limits_{\sigma = 0}^{q - 1} \frac{\widehat{T}_B }{1 - \exp (2 \pi \I \Delta x  \, B)}  \exp(- 2 \pi \I  \Delta p \, B d_1),
\ee
$\, B = q^{-s}(D + \sigma/q).$
\subsection{Binary system with arbitrary digit}
Hence, for the binary system (q = 2) with arbitrary digits $\{d_{1}, d_{2} = d_{1}+1 \}$ we get the following expression:
\begin{eqnarray} \label{gen_f}
	\hat{p}_r  = \frac{d_{1} + d_{2}}{2} \, \cdot \, \hat{1} - \Delta p \, 2^{-r} \; \sum\limits_{D \in \mathds{Z}_{2^{r}/\Delta p}} \frac{\widehat{T}_{-A}}{1 - \exp(2\pi \I \Delta p A)} \exp(- 2 \pi \I \Delta p A d_{1}) =  \\ 
=  \frac{d_{1} + d_{2}}{2} \, \cdot \, \hat{1} + \Delta p \, 2^{-r} \; \sum\limits_{D \in \mathds{Z}_{2^{r}/\Delta p}} \frac{\widehat{T}_{-A}}{2\I \sin(\pi \I \Delta p A)} \exp(- \pi \I \Delta p A \left(2d_{1} + 1\right)) \nonumber.
\end{eqnarray}
We will later use the case of such binary systems to demonstrate, that the shifted lattices tend to be ``natural`` in some cases.
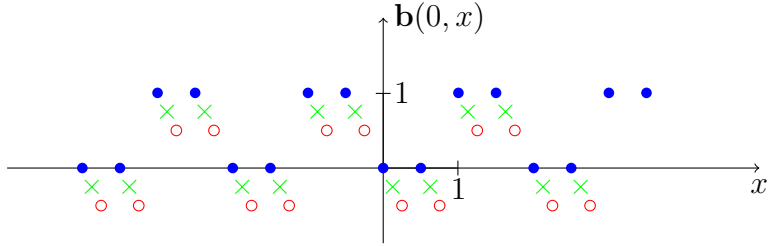
\begin{figure}[h]
\centering
\begin{tikzpicture}
\draw [->] (-5,0) -- (5,0) node [below] {$x$};
\draw [-|] (0,0) -- (1,0) node [below] {1};
\draw [-|] (0,0) -- (0,1) node [right] {1};
\draw [->] (0,-1) -- (0,2) node [right] {$\mathbf{b}(0,x)$};

\fill[blue] (-4,0) circle (2pt);
\fill[blue] (-3.5,0) circle (2pt);

\fill[blue] (-3,1) circle (2pt);
\fill[blue] (-2.5,1) circle (2pt);

\fill[blue] (-2,0) circle (2pt);
\fill[blue] (-1.5,0) circle (2pt);

\fill[blue] (-1,1) circle (2pt);
\fill[blue] (-0.5,1) circle (2pt);

\fill[blue] (0,0) circle (2pt);
\fill[blue] (0.5,0) circle (2pt);

\fill[blue] (1,1) circle (2pt);
\fill[blue] (1.5,1) circle (2pt);

\fill[blue] (2,0) circle (2pt);
\fill[blue] (2.5,0) circle (2pt);

\fill[blue] (3,1) circle (2pt);
\fill[blue] (3.5,1) circle (2pt);

\draw[red] (-3.75,-0.5) circle (2pt);
\draw[red] (-3.25,-0.5) circle (2pt);

\draw[red] (-2.75,0.5) circle (2pt);
\draw[red] (-2.25,0.5) circle (2pt);

\draw[red] (-1.75,-0.5) circle (2pt);
\draw[red] (-1.25,-0.5) circle (2pt);

\draw[red] (-0.75,0.5) circle (2pt);
\draw[red] (-0.25,0.5) circle (2pt);

\draw[red] (0.25,-0.5) circle (2pt);
\draw[red] (0.75,-0.5) circle (2pt);

\draw[red] (1.25,0.5) circle (2pt);
\draw[red] (1.75,0.5) circle (2pt);

\draw[red] (2.25,-0.5) circle (2pt);
\draw[red] (2.75,-0.5) circle (2pt);

\fill[green] (-3.75 - 0.125,-0.5 + 0.25) node[cross=3pt,green]{};
\fill[green] (-3.25 - 0.125,-0.5 + 0.25) node[cross=3pt,green]{};

\fill[green] (-2.75 - 0.125,0.5 + 0.25) node[cross=3pt,green]{};
\fill[green] (-2.25 - 0.125,0.5 + 0.25) node[cross=3pt,green]{};

\fill[green] (-1.75 - 0.125,-0.5 + 0.25) node[cross=3pt,green]{};
\fill[green] (-1.25 - 0.125,-0.5 + 0.25) node[cross=3pt,green]{};

\fill[green] (-0.75 - 0.125,0.5 + 0.25) node[cross=3pt,green]{};
\fill[green] (-0.25 - 0.125,0.5 + 0.25) node[cross=3pt,green]{};

\fill[green] (0.25 - 0.125,-0.5 + 0.25) node[cross=3pt,green]{};
\fill[green] (0.75 - 0.125,-0.5 + 0.25) node[cross=3pt,green]{};

\fill[green] (1.25 - 0.125,0.5 + 0.25) node[cross=3pt,green]{};
\fill[green] (1.75 - 0.125,0.5 + 0.25) node[cross=3pt,green]{};

\fill[green] (2.25 - 0.125,-0.5 + 0.25) node[cross=3pt,green]{};
\fill[green] (2.75 - 0.125,-0.5 + 0.25) node[cross=3pt,green]{};


\end{tikzpicture}
\caption{Plot of the value of the binary digit number $0$, on a lattice ($n_{-} = 1$) for an ``symmetric system'' ($d_1 = -0.5$) -- red circles,  $d_{1} = -0.25$ -- green crosses, and ''non-symmetric system'' $d_1 = 0$ -- blue circles.}
\label{systems}
\end{figure}

\subsection{Ternary symmetric system}
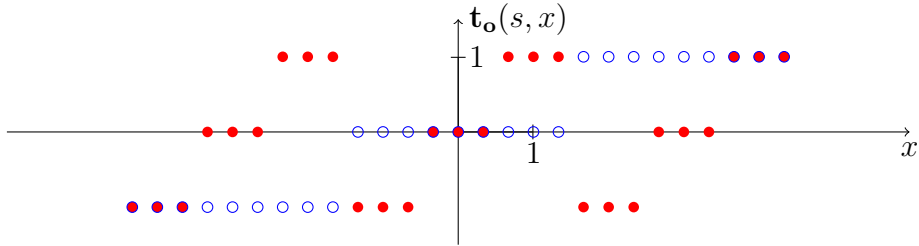
\begin{figure}[h]
    \centering
\begin{tikzpicture}
\draw [->] (-6,0) -- (6,0) node [below] {$x$};
\draw [-|] (0,0) -- (1,0) node [below] {1};
\draw [-|] (0,0) -- (0,1) node [right] {1};
\draw [->] (0,-1.5) -- (0,1.5) node [right] {$\mathbf{t_o}(s,x)$};

\fill[red] (-4.333,-1) circle (2pt);
\fill[red] (-4,-1) circle (2pt);
\fill[red] (-3.666,-1) circle (2pt);

\fill[red] (-3.333,0) circle (2pt);
\fill[red] (-3,0) circle (2pt);
\fill[red] (-2.666,0) circle (2pt);

\fill[red] (-2.333,1) circle (2pt);
\fill[red] (-2,1) circle (2pt);
\fill[red] (-1.666,1) circle (2pt);

\fill[red] (-0.666,-1) circle (2pt);
\fill[red] (-1,-1) circle (2pt);
\fill[red] (-1.333,-1) circle (2pt);

\fill[red] (-0.333,0) circle (2pt);
\fill[red] (0,0) circle (2pt);
\fill[red] (0.333,0) circle (2pt);

\fill[red] (0.666,1) circle (2pt);
\fill[red] (1,1) circle (2pt);
\fill[red] (1.333,1) circle (2pt);

\fill[red] (1.666,-1) circle (2pt);
\fill[red] (2,-1) circle (2pt);
\fill[red] (2.333,-1) circle (2pt);

\fill[red] (2.666,0) circle (2pt);
\fill[red] (3,0) circle (2pt);
\fill[red] (3.333,0) circle (2pt);

\fill[red] (3.666,1) circle (2pt);
\fill[red] (4,1) circle (2pt);
\fill[red] (4.333,1) circle (2pt);

\draw [color=blue] (-1.666,-1) circle (2pt);
\draw [color=blue] (-2,-1) circle (2pt);
\draw [color=blue] (-2.333,-1) circle (2pt);
\draw [color=blue] (-2.666,-1) circle (2pt);
\draw [color=blue] (-3,-1) circle (2pt);
\draw [color=blue] (-3.333,-1) circle (2pt);
\draw [color=blue] (-3.666,-1) circle (2pt);
\draw [color=blue] (-4,-1) circle (2pt);
\draw [color=blue] (-4.333,-1) circle (2pt);

\draw [color=blue] (-1.333,0) circle (2pt);
\draw [color=blue] (-1,0) circle (2pt);
\draw [color=blue] (-0.666,0) circle (2pt);
\draw [color=blue] (-0.333,0) circle (2pt);
\draw [color=blue] (0,0) circle (2pt);
\draw [color=blue] (0.333,0) circle (2pt);
\draw [color=blue] (0.666,0) circle (2pt);
\draw [color=blue] (1,0) circle (2pt);
\draw [color=blue] (1.333,0) circle (2pt);

\draw [color=blue] (1.666,1) circle (2pt);
\draw [color=blue] (2,1) circle (2pt);
\draw [color=blue] (2.333,1) circle (2pt);
\draw [color=blue] (2.666,1) circle (2pt);
\draw [color=blue] (3,1) circle (2pt);
\draw [color=blue] (3.333,1) circle (2pt);
\draw [color=blue] (3.666,1) circle (2pt);
\draw [color=blue] (4,1) circle (2pt);
\draw [color=blue] (4.333,1) circle (2pt);

\end{tikzpicture}
\caption{A plot of the value of the ternary digit number $s$ for a finite lattice ($n_{-} = 1$), $s=0$ -- red filled circles, $s=1$ -- blue circles.}
\end{figure}
The other case, which was previously discussed in \cite{ter-bin}, where shifted lattice emerged implicitly is the ternary system with digits $\{-1,0,1\}$, also called \textit{ternary symmetric system}. For this system we can obtain the following expression (explicitly or using the general formula):
\begin{equation}
    \hat{p}_r =  \Delta p\, 3^{-r}\sum\limits_{D \in \mathds{Z}_{3^r/\Delta p}} \sum\limits_{\sigma = 1}^{2}  \frac{(-1)^{D + \sigma}}{2 \I \sin\left(\pi\,\Delta p \,A\right)}\widehat{T}_{-A},\quad 
    A=3^{-r}(D + \sigma/3).
\end{equation}

\subsection{Renormalizations of infinite and finite sums}
Usually the renormalization procedure consists of two steps -- the introduction of a cutoff, and the removal of the cutoff. In our case the role of the cutoff plays the lattice itself and the parameter of the cutoff is the parameter of the lattice. In this paper the physical applications of such renormalization are not being discussed. The development of this idea will be presented in \cite{numth_renorm}.
\subsubsection{Motivation for renormalizing infinite sums}
Let's start with the plot of the digit number $0$ for the binary non-symmetric system:
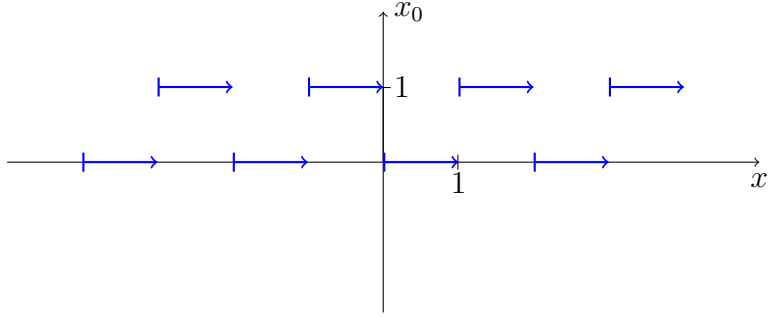
\begin{figure}[h]
\centering
\begin{tikzpicture}
\draw [->] (-5,0) -- (5,0) node [below] {$x$};
\draw [-|] (0,0) -- (1,0) node [below] {1};
\draw [-|] (0,0) -- (0,1) node [right] {1};
\draw [->] (0,-2) -- (0,2) node [right] {$x_0$};

\draw[|->,thick,blue] (0,0)--(1,0);
\draw[|->,thick,blue] (1,1)--(2,1);
\draw[|->,thick,blue] (2,0)--(3,0);
\draw[|->,thick,blue] (3,1)--(4,1);

\draw[|->,thick,blue] (-4,0)--(-3,0);
\draw[|->,thick,blue] (-3,1)--(-2,1);
\draw[|->,thick,blue] (-2,0)--(-1,0);
\draw[|->,thick,blue] (-1,1)--(0,1);
\end{tikzpicture}
\caption{A plot of the value of the binary digit number $0$ for binary non-symmetric system on the line}
\end{figure}
The digit number $s$ can be obtained by scaling the entire plot over the $x$-axis on multiplier $2^{s}$. Hence, it is easy to see that for any negative number $x$ there is number $n$, for which all the digits with index greater then $n$ will be equal to 1, which means that the sum
\be
 x \sim \sum\limits_{s = -\infty}^{\infty} x_{s} 2^{s},
\ee
is diverging. The problem becomes more explicit when we turn to the binary symmetric system:
\begin{figure}[h]
\centering
\begin{tikzpicture}
\draw [->] (-5,0) -- (5,0) node [below] {$x$};
\draw [-|] (0,0) -- (1,0) node [below] {1};
\draw [-|] (0,0) -- (0,1) node [right] {1};
\draw [->] (0,-2) -- (0,2) node [right] {$x_0$};

\draw[|->,thick,red] (0,-0.5)--(1,-0.5);
\draw[|->,thick,red] (1,0.5)--(2,0.5);
\draw[|->,thick,red] (2,-0.5)--(3,-0.5);
\draw[|->,thick,red] (3,0.5)--(4,0.5);

\draw[|->,thick,red] (-4,-0.5)--(-3,-0.5);
\draw[|->,thick,red] (-3,0.5)--(-2,0.5);
\draw[|->,thick,red] (-2,-0.5)--(-1,-0.5);
\draw[|->,thick,red] (-1,0.5)--(0,0.5);
\end{tikzpicture}
\caption{A plot of the value of the binary digit number $0$ for binary symmetric system on the line}
\end{figure}
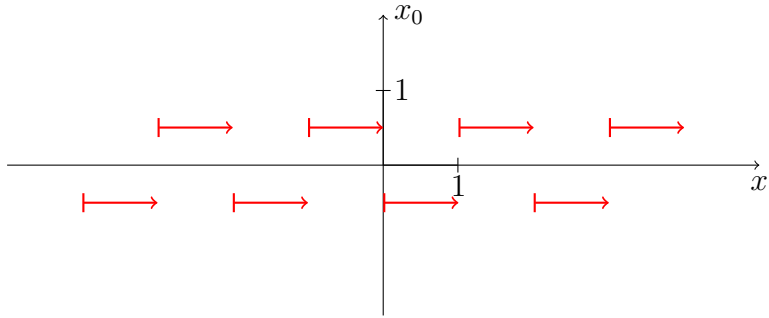

in this case we obtain the divergence for both negative and positive numbers.

\subsubsection{Renormalization of infinite sums}
To solve the emerging problem, let's consider the arbitrary positional base-$q$ system with digits $x_s$. As we have seen, for such a system the row
\be\label{geom}
\sum\limits_{s = -\infty}^{\infty}x_s q^s,
\ee
does not converge in general case. For this case we can introduce the sum ``with prime``, which is determined in the following way:
\be
\sum\limits_{s = 0}^{\infty \, \prime}q^s = \frac{1}{1-q} \, ,
\ee
which results in application of the formula for the sum of converging geometrical progression over the field of its application. Such ``renormalization`` results in the expression
\be
\sum\limits_{s \in \mathds{Z}}^{\quad \prime}q^s = 0,
\ee
which is, generally speaking, the alternative definition of this renormalization.

It is possible in some situations to consider the row (\ref{geom}) in $q$-adic sence, as, for instance in \cite{p-adic}, it will result in the convergence of the digits with infinite number of non-zero digits after the point, but the divergence of numbers with infinite number of non-zero digits before the point in the other hand.

The alternative way to renormalize the diverging sum, inspired by the examples considered above, can be represented by the following formal calculation:
\be
x = \frac{qx - x}{q-1} = \frac{1}{q - 1}\sum\limits_{s \in \mathds{Z}} \, (x_{s - 1} - x_{s})q^s.
\ee
Obviously, both ways to renormalize the sum fix the problems, which occur when we try to write the number as the row of digits in the positional system. Now we are ready to turn back to the lattice.
\subsubsection{Renormalization on the lattice}
As far as the lattice is finite, we can not obtain the divergent sums on it. Therefore, the purpose of the renormalization is no longer to avoid the divergence, but to \textit{change the representation} of $\mathds{Z}_{N}$ from $\{0,1,\hdots,N-1\}$ to, for instance, $\{-k, -k + 1, \hdots , -k + N - 1\}$.

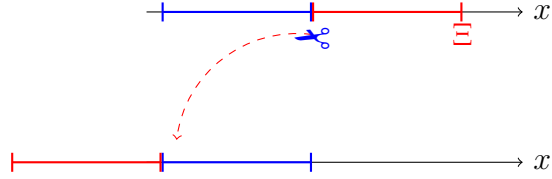
\begin{figure}[h]
\centering
\begin{tikzpicture}
\coordinate (A) at (1.2,-0.3);
\coordinate (B) at (-0.6,-1.7);

\draw[->, red, dashed] (A) to [bend right=45] (B);

\draw[->](-1,0) -- (4,0) node [right] {$x$};
\draw[|-|, blue, thick](-0.8,0) -- (1.2,0) node [below]{\ScissorLeftBrokenTop};

\draw[|-|, red, thick](1.2,0) -- (3.2,0) node [below]{$\Xi$};

\draw[->](-1,-2) --(4,-2) node [right] {$x$};

\draw[|-|, blue, thick](-0.8,-2) -- (1.2,-2);
\draw[|-|, red, thick](-2.8,-2) -- (-0.8,-2);
\end{tikzpicture}
\end{figure}
To demonstrate the concept more precise, let's consider two examples. Let us start with the binary system. For binary non-symmetric system, we obtain the following expressions:
\be
x = \sum\limits_{s = -n_{-}}^{n_{+} - 1~\prime}x_s 2^s = 
	\sum\limits_{s = -n_{-}}^{n_{+} - 1} (x_{s-1} - x_s)2^{s}, \quad x_{-n_{-} - 1} = 0,
\ee
such renormalization is equal to change of the last digit  $x'_{n_+-1}=-x_{n_+-1}$:
\be
\sum\limits_{s = -n_{-}}^{n_{+} - 1~\prime \prime}x_s 2^s
= \sum\limits_{s = -n_{-}}^{n_{+} - 2}x_s 2^s +x'_{n_+-1}2^{n_+-1}.
\ee
and is linear in accordance to the binary digits $x_s$.

It is noteworthy, that in the ternary case the linear renormalization does not work. Indeed, the linear renormalization is described by the expression:
\be
    x^{\prime} = \sum\limits_{s = -n_{-}}^{n_{+} - 1~\prime}\mathbf{t_{ns}}(s,x) 3^s = 
    \frac{1}{2}\sum\limits_{s = -n_{-}}^{n_{+} - 1} (\mathbf{t_{ns}}(s - 1, x) - \mathbf{t_{ns}}(s,x))3^{s}, \quad \mathbf{t_{ns}}(-n_{-} - 1, x) = 0.
\ee
Maximal ternary number is renormalized in the right way:
\be
    \sum\limits_{s = -n_{-}}^{n_{+} - 1 ~\prime}2\cdot 3^s = \frac{1}{2} (-2\cdot 3^{-n_{-}}) =-\Delta x,
\ee
but the half is not the node of the lattice:
\be
\sum\limits_{s = -n_{-}}^{n_{+} - 1 ~\prime} 3^s =-\frac{\Delta x}{2}.
\ee

The alternative (and properly working) way to obtain the lattice
\\ $\{-3^{n-1}\Delta x,\dots,-\Delta x,0,\Delta x,2\Delta x,\dots,(3^n-3^{n-1}-1)\Delta x\}$ from the initial one ($\{0,\Delta x,2\Delta x,3\Delta x,\dots, (3^n-1)\Delta x\}$) is to substract $3^{n_+}$ from the last $3^{n-1}$ nodes. Then we obtain:
\bea
  \mathbf{t^{\prime\prime}_{ns}}(n_+-1,x)&=&\left\{\begin{array}{cc}
   0,&\mathbf{t_{ns}}(n_+-1,x)=0,\\
   1,&\mathbf{t_{ns}}(n_+-1,x)=1,\\
   -1,&\mathbf{t_{ns}}(n_+-1,x)=2
  \end{array}\right.=\\ \nonumber
  &=&\mathbf{t_{ns}}(n_+-1,x)-\frac32(\mathbf{t_{ns}}(n_+-1,x)-1)\mathbf{t_{ns}}(n_+-1,x).
\eea
\begin{equation}
x^{\prime\prime} = \sum\limits_{s = -n_{-}}^{n_{+} - 1~\prime\prime}\mathbf{t_{ns}}(s,x) 3^s 
= \sum\limits_{s = -n_{-}}^{n_{+} - 2}\mathbf{t_{ns}}(s,x)\, 3^s +\mathbf{t^{\prime\prime}_{ns}}(n_+-1,x)\, 3^{n_+-1}.
\end{equation}

Such renormalization works, but is non-linear  to the ternary digits $x_s$.

\subsubsection{Digit on the line}
After defining the infinite sums of digits, we can discuss the limit of $n \rightarrow +\infty$, $\Delta p \rightarrow 0$. In this case we obtain the following expression for the momentum digit on the line:
\be
\hat{p}_{r} = \frac{d_1 + q + 1}{2} \; \hat{1} + q^{-r} \sum\limits_{D \in \mathds{Z}_{q^r/\Delta p} } \sum\limits_{\sigma = 0}^{q - 1}  \frac{\widehat{T}_A}{2 \pi \I A}, \; A = q^{-r}(D + \sigma/q).
\ee
This limit corresponds to the infinite number of digits before the point for the coordinate and after the point for the momentum. Hence, we have the infinite coordinate lattice, while the momentum lattice is periodical (and if we additionally consider the limit $\Delta x \rightarrow 0$ -- then we obtain the real lines for both coordinate and momentum). It is noteworthy, that after transition to the limit the sum is taken over both positive and negative integers. On the finite periodical lattice the concept of positive and negative number was not defined.

\section{Commutation relations}
On the line occurs the well-known canonical commutation relation between the coordinate and momentum. On the lattice it is impossible to define coordinate and momentum operators such, that they will satisfy the canonical commutation relation.
\subsection{Digit-digit commutator}
It is easy to derive the commutation relation between an arbitrary function of the coordinate $f(\hat{x})$ and the shift operator $\widehat{T}_{A}$:
\begin{equation}
    [f(\widehat{x}), \widehat{T}_{A}]\psi(x) = (f(\widehat{x}) - f(\widehat{x} + A))\widehat{T}_{A}\psi(x).
\end{equation}
Hence,
\begin{equation}
    [f(\widehat{x}), \widehat{T}_{A}] = (f(\widehat{x}) - f(\widehat{x} + A))\widehat{T}_{A}.
\end{equation}
Using notation $d(s, \hat{x}) = \hat{x}_s$, we obtain the form of the commutators of the digits of the coordinate and of the momentum on the lattice:
\be
       [\hat{x}_s,\hat{p}_r] =    
       - \Delta p \cdot q^{-r} \,
       \sum\limits_{D \in \mathds{Z}_{q^r/\Delta p}}
        \sum\limits_{\sigma = 1}^{q-1}
        \frac{\{d(s,\hat{x}) - d(s, \hat{x} - q^{-r}(D + 1/q)\} \widehat{T}_{-A}}{1 - \exp(2 \pi \I \Delta p A)} \exp(-2\pi \I A \Delta p d_1).
 \ee

As in \cite{bin-ivanov} we have that the requirement for commutation is following:
\be
	-r - s - 2 \geq 0, \quad s + r \leq -2.
\ee
Hence, the fractional part of the momentum commutes with the fractional part of the coordinate, the lowest digit of momentum does not commute only with the highest digit of the coordinate, and lowest digit of the coordinate does not commute only with the highest digit of the momentum. The fractional parts of the coordinate and momentum can be considered as the full set of observables for one-dimensional  motion.

\subsection{Coordinate - digit commutator}
Because $\hat x = \sum\limits_{s = -n_{-}}^{n_{+} - 1} q^s d(s, \hat x)$, we obtain the following commutator of coordinate and the digit of momentum:
\be
	[\hat x, d(r, \hat{p})] = - \frac{1}{q^r} \sum\limits_{D \in \mathbb{Z}_{q^r/\Delta p}} 
	 \sum\limits_{\sigma = 1}^{q-1}
	\frac{\Delta p \cdot q^{-r} (D + 1/q)}{1 - \exp(2 \pi \I \Delta p A)} \widehat{T}_{-A} \exp(-2\pi \I A \Delta p d_1).
\ee
\subsection{Coordinate-momentum commutator}
Similarly, we consider $\hat p = \sum\limits_{r = n_{+}}^{n_{-} - 1} q^r d(r, \hat p)$ and hence obtain the following commutator:
\be
	[\hat x, \hat p] = -\sum\limits_{r = -n_{+}}^{n_- - 1}
	\sum\limits_{D \in \mathbb{Z}_{q^r/\Delta p}} 
	 \sum\limits_{\sigma = 1}^{q-1}
	 \frac{\Delta p \cdot q^{-r} (D + 1/q)}{1 - \exp(2 \pi \I \Delta p A)} \widehat{T}_{-A} \exp(-2\pi \I A \Delta p d_1).
\ee

\subsection{Commutators on the line}
On the line the commutators become independent of $d_1$ (the phase shift becomes infinitely small). The formulas have the following view:
\be
	[d(s,\hat x), d(p, \hat r)] = \sum\limits_{D \in \mathbb{Z}} 
	 \sum\limits_{\sigma = 1}^{q-1}
	 \frac{d(s, \hat x) - d(s, \hat x - q^{-r}(D + \sigma/q)}{2\pi \I (D + \sigma/q)} \widehat{T}_{-A},
\ee
\be
	[\hat x, d(r, \hat p)] = \frac{1}{q^{r}} \frac{1}{2\pi \I} \sum\limits_{D \in \mathbb{Z}} 
	 \sum\limits_{\sigma = 1}^{q-1}
	 \widehat{T}_{-A},
\ee
\be\label{commutator}
	[\hat x, \hat p] = \frac{1}{2 \pi \I}\sum\limits_{r \in \mathbb{Z}} \sum\limits_{D \in \mathbb{Z}} 
	 \sum\limits_{\sigma = 1}^{q-1}
	 \widehat{T}_{-A}.
\ee

\subsection{Renormalization of the commutator on the line}
We obtain a formal decomposition of the commutator in the sum of the shift operators, as $\hbar = 1/2\pi$, we can rewrite (\ref{commutator}) in the following way:
\be\label{commutator_2}
[\hat x, \hat p] = - \I \hbar \sum\limits_{r \in \mathbb{Z}} \sum\limits_{D \in \mathbb{Z}} 
	 \sum\limits_{\sigma = 1}^{q-1}
	 \widehat{T}_{-A}.
\ee

Let $\mathds{A}$ be a set of numbers whose $q$-nary expansion contains a finite number of nonzero factors with negative powers of $q$ (a finite number of significant $q$-nary digit after $q$-nary point); $\mathbb{A}$ is a group under the summation operation.
Then the set of values of the shifts along which summation occurs has the form $\mathbb{A} / 0$. Given that $\widehat{T}_0 = \hat{1}$, we obtain:
\begin{equation}
    [\hat{x}, \hat{p}] = -\I \hbar \sum\limits_{A \in \mathds{A} \setminus \{0\}}\widehat{T}_{A} = -\I \hbar \left(\sum\limits_{A \in \mathds{A}}\widehat{T}_{A} - \hat{1} \right) = \I \hbar \hat{1} - \I \hbar \sum\limits_{A \in \mathds{A}}\widehat{T}_{A}.
\end{equation}
We know that for the particle coordinate and momentum on the line, there is the canonical commutation relation $[\hat{x}, \hat{p}] = i\hbar \hat{1}$. We thus obtain the renormalization
\begin{equation}
    \sum\limits_{A \in \mathds{A}}\widehat{T}_{A} = \sum\limits_{A \in \mathds{A}} \E ^{2\pi \I A \hat{p}} = 0.
\end{equation}
This renormalization is similar to the formal equality $\int\limits_{\mathds{R}}\E ^{2\pi \I x p}dx = 0$ for all $p \neq 0$ arising in the Fourier transforms.

\section{Physical motivation for shifted lattice}
\subsection{Boundary conditions}
Let us denote the binary digit number $r$ of the momentum $p$ in the system with digits $\{d_1, d_1 + 1\}$ as $b(r,p,d_1)$. From the equation for the digit in the binary system \eqref{gen_f} we can see, that
\begin{equation} \label{phase_0}
b(r,p,d_1) =  d_1+ b(r,p,0) \cdot \E^{-2\pi\I\Delta p A d_1}.
\end{equation}
Therefore, we can mention that the shift of momentum lattice induces the change of boundary conditions for the coordinate. The boundary conditions for the momentum with digits $\{ d_{1}, d_{1} + 1\}$ can be written in the following form:
\begin{equation} \label{phase}
	\Psi(x + 2^{n_{+}}) = \E^{i\phi} \cdot \Psi(x), \, \text{where} ~ \phi = -2\pi d_{1}.
\end{equation}

\begin{figure}[h!]
\begin{center}
\begin{tikzpicture}
\draw[blue,thick] (0,0) circle (2);
\draw[black,fill=green,thick] (0,2) circle (0.1) node [above] {$\E^{\I \phi}  |\psi \rangle \leftarrow 3d_1 + 4=3d_1 \rightarrow |\psi \rangle$};
\draw[black,fill=red,thick] ({2)},{0}) circle (0.1) node [below right] {$3d_1 + 1$};
\draw[black,fill=red,thick] ({0},{-2}) circle (0.1) node [below left] {$3d_1 + 2$};
\draw[black,fill=red,thick] ({-2},{0}) circle (0.1) node [below left] {$3d_1 + 3$};

\end{tikzpicture}
\caption{The rotation over the lattice period gives an additional phase to the wavefunction.}
\end{center}
\end{figure}
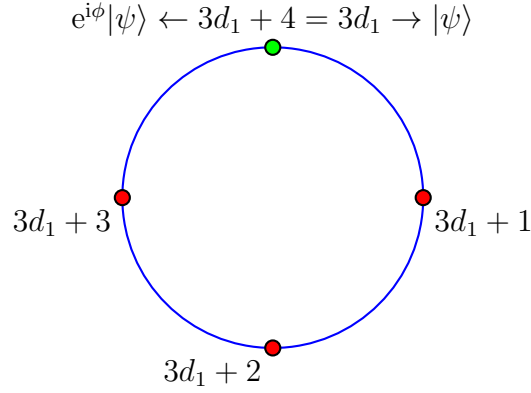
\subsection{Aharonov–Bohm effect}
In \cite{shift} was established a relation between flux of the magnetic field passing through the torus, shift of the generalized momentum and boundary conditions of the wavefunction. We analogically can consider the ring, through which there is non-zero flux of the magnetic field $\Phi$. 
\\
The stationary state of a charged particle with the charge $e$ and mass $m$ on this ring of length $a$ is the eigenfunction of the operator
\begin{equation}
	\hat{H}_0 = -\frac{\hbar^2}{2m}\left( \frac{\partial}{\partial x} - \I\frac{e}{c\hbar}A_{x}(x)\right)^2,
\end{equation}
where $\mathbf{A}(x)$ is a vector potential.
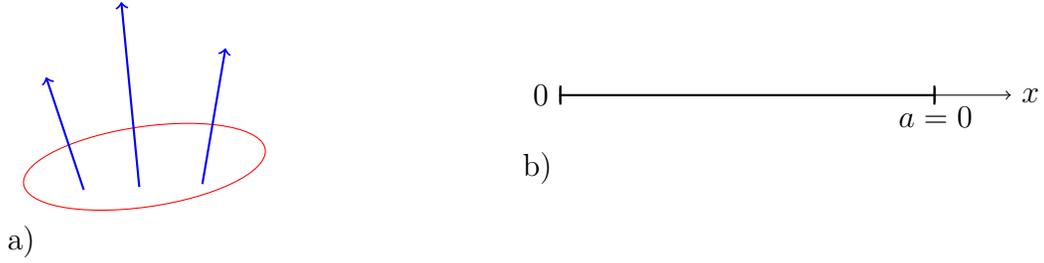
\begin{figure}[h!]
\begin{minipage}[h]{0.49\linewidth}
\begin{tikzpicture}
\draw[red] (1.5, 1, 1.5) [y={(0,0,1)}] circle (1.5);
\draw [->, thick, blue] (0,0,-0.3) -- (0,2,1);
\draw [->, thick, blue] (1.5,0,-0.5) -- (2,2,0);
\draw [->, thick, blue] (0.7,0,-0.4) -- (1,3,1);
\end{tikzpicture}
\\ a)
\end{minipage}
\begin{minipage}[h]{0.49\linewidth}
\begin{tikzpicture}
\draw[|-|, thick](0,0) --(5,0) node [below] {$a = 0$};
\draw[|-|](5,0) --(0,0) node [left] {$0$};
\draw[->](0,0) --(6,0) node [right] {$x$};
\end{tikzpicture}
\\ b)
\end{minipage}
\caption{a) Ring in the magnetic field, b) Coordinates on the ring}
\end{figure}
\\
Function $\Psi(x)$ satisfies to periodic boundary conditions:
\begin{equation}
\Psi(0) = \Psi(a),
\end{equation} 
\begin{equation}
\Psi^{\prime}(0) = \Psi^{\prime}(a),
\end{equation} 
After gauge transformation we can nullify $\mathbf{A}$ on the ring. This will lead to the discontinuity of wavefunction at the cuts. It is also clear that $|\Psi(x)|^{2}$ should not change. Then the problem is simplified to the problem of finding eigenfunctions $\psi(x)$ of an operator
\bea \label{H_1}
	\hat{H}_1 = -\frac{\hbar^2}{2m}\left(\frac{\partial}{\partial x} \right)^2,
\eea
with the phase shift boundary conditions
\bea \label{bound_1}
	\psi(0) = \E^{\I \phi} \psi(a), \quad \psi^{\prime}(0) =  \E^{\I \phi} \psi^{\prime}(a).
\eea
The gauge transformation, nullifying the vector potential on the ring can be written in the following form:
\bea \label{gauge}
\mathbf{A}^{\prime}(x) = \mathbf{A} - \nabla f(x) = 0.
\eea
Because we want the probability $|\Psi(x)|^{2}$ not to be changed by gauge transformation \eqref{gauge}, the new wavefunction $\psi(x)$ differs from the old one $\Psi(x)$ only in a phase factor:
\bea \label{phase_2}
\psi(x) = \Psi(x)\exp\left(
	-\frac{\I e}{\hbar c} f(x)
\right).
\eea
For Hamiltonian \eqref{H_1} we can easily find the eigenfunctions, fitting the boundary conditions \eqref{bound_1}:
\bea
	\psi(x) = \E^{\I k x}, \, \text{where}~ k = \frac{2\pi n}{a} - \frac{\phi}{a}.
\eea
Now let's consider the relationship between flux of magnetic field $\Phi$ and phase shift $\phi$. 
\bea
	\Phi =  f(a) - f(0) = \underbrace{\frac{\hbar c}{e}}_{\frac{\Phi_0}{2\pi}} \phi.
\eea
From this example we can see, that in a real physical system a shift of (generalized) momentum induces the shift of wavefunction itself (as in \eqref{phase_0}) and the phase factor in boundary conditions (as in \eqref{phase}). In such cases use of the shifted lattice for momentum is quite natural.

\section{Conclusion}
This paper generalizes the results of previously published works \cite{bin-ivanov} and \cite{ter-bin} for the case of the $q$-base numeral systems. The only restriction on the digits of considered positional systems is that the distance between adjacent digits is $1$.

The possible application for the obtained expansions is not only the quantum computations, but also, for instance, the solution of partial differential equations on classical computer. It worth mentioning that since we define the momentum operator as the shift generator, it appears to be non-local (see appendix A), which helps to better describe the symmetries of considered theory.

The natural emergence of renormalizations is particularly interesting. Renormalizing infinite and finite quantities (on the lattice) allows finding the renormalization numerically by passing from the lattice to the limit of a continuous quantity.
Time and energy can also be considered as a coordinate and a momentum \cite{dyn_time}, which allows applying the same renormalization methods to them. Renormalizations in this context are probably related to the quantum theory of measurements (see \cite{book}, \cite{measurement} and the references therein for the quantum theory of measurements).

The digital representation assumes that the coordinate and the momentum are not observed in the experiment. Instead, individual digits of the coordinate and the momentum are directly observed. The measurement of a binary digit of the spatial coordinate corresponds to the passage/nonpassage of a particle through a diffraction grating.
It's quite interestingly, that the system with non-integer digits appears to be in some particular cases (e.g. for non-trivial boundary conditions) more ``natural`` then the common one. It is also noteworthy that the lattices of such systems induce the constructions, which are quite similar to the windings of torus (see appendix B), which may be applicable to ergodicity.
\section{Acknowledgments}
The authors thank the participants of the Conference Phystech-Quant 2020 (Moscow Institute of Physics and Technology, 2020), of the section of theoretical physics of the 62-nd, 63-rd and 64-th Science Conferences of MIPT (2019, 2020 and 2021 respectively), 
the seminar of the Department of mathematical physics (Steklov mathematical institute)
and the seminar of the Laboratory of infinite-dimensional analysis and mathematical physics (Moscow State Univercity). Separately, the authors thank I. V. Volovich, Z. V. Khaidukov, V. A. Dudchenko, V. V. Naumov, N. N. Shamarov, V. Zh. Sakbayev and other colleagues for useful discussion.

\section*{AppendixA}
Let us consider the matrices of the momentum digits and operators in the cases, when they are compact enough to be placed on paper.

Everywhere in this section, $\Delta x = 1, \, x \in \mathds{Z}_{N} \in \{0,1, ... ,N - 1  \}$, coordinates and momenta are numbered by ternary numbers, which are marked with a lower index 3, and $\Delta p = 3^{-n} = 1/N$.
\subsection*{A1. The case $n = 1$ and $N = 3^1 = 3$. Symmetric system.} In this case, we have 
$$\hat x = \hat x_0 = \left(
\begin{array}{ccc}
     +1&0&0  \\
     0&0&0  \\
     0&0&-1
\end{array}
\right) = \hat{s}_{z}
$$

$$
\hat p_{-1} =\frac{1}{\sqrt{3}} \left(
\begin{array}{ccc}
    0 & \I & -\I \\
    -\I & 0 & \I \\
    \I & -\I & 0
\end{array}
\right) = \frac{1}{\sqrt{3}} \left(\sqrt{2}\hat{s}_{y} + 2\hat{s}_{y}\hat{s}_{x} + \I \hat{s}_{z} \right)
$$

\subsection*{A2. The case $n = 1$ and $N = 3^1 = 3$. Non-symmetric system.}
Here we obtain following results:
$$\hat x = \hat x_0 = \left(
\begin{array}{ccc}
     2&0&0  \\

     0&1&0  \\
     0&0&0
\end{array}
\right) = \hat{1} + \hat{s}_z; \quad
\hat p_{-1} =\frac{1}{6} \left(
\begin{array}{ccc}
     6& 1 - \sqrt{3}\,\I& 1 + \sqrt{3}\,\I  \\
     1 + \sqrt{3}\,\I& 6 & 1 - \sqrt{3}\,\I \\
     1 - \sqrt{3}\,\I& 1 + \sqrt{3}\,\I& 6
\end{array}
\right) =
$$
$$=\hat{1} +  \frac{\sqrt{2}}{6}\hat{s}_{x} + \frac{1}{\sqrt{6}}\hat{s}_{y} - \frac{\sqrt{3}}{6}(2\hat{s}_{y}\hat{s}_{x} + \I \hat{s}_{z})
$$
\subsection*{A3. The case $n = 2$ and $N = 3^2 = 9$. Non - symmetric system.}
$$x = x_0 + 3\cdot x_1=\text{diag}(8;7;6;5;4;3;2;1;0),$$
$$x_0=\text{diag}(2;1;0;2;1;0;2;1;0),$$
$$x_1=\text{diag}(2;2;2;1;1;1;0;0;0).$$
 
Let's denote:
$$
    E_{n} =  \frac{1}{\exp\left(\frac{-2 \pi \I n}{9}\right) - 1},
    $$
    Then:
$$
\hat{p}_{-1} = \frac{1}{3}\left(
\begin{array}{ccccccccc}
     3&E_{8} &E_{7} & 0& E_{5}& E_{4}&0 &E_{2} & E_{1}\\
     E_{1} & 3 &E_{8} &E_{7} & 0& E_{5}& E_{4}&0 &E_{2}\\
     E_{2} &E_{1} & 3 &E_{8} &E_{7} & 0& E_{5}& E_{4}&0\\
     0&E_{2} &E_{1} & 3 &E_{8} &E_{7} & 0& E_{5}& E_{4}\\
     E_{4} &0&E_{2} &E_{1} & 3 &E_{8} &E_{7} & 0& E_{5}\\
     E_{5} &E_{4} &0&E_{2} &E_{1} & 3 &E_{8} &E_{7} & 0\\
     0 &E_{5} &E_{4} &0&E_{2} &E_{1} & 3 &E_{8} &E_{7} \\
     E_{7} &0 &E_{5} &E_{4} &0&E_{2} &E_{1} & 3 &E_{8} \\
     E_{8} & E_{7} &0 &E_{5} &E_{4} &0&E_{2} &E_{1} & 3
\end{array}
\right).
$$
   $$
   \hat{p}_{-2} = \left(
   \begin{array}{ccccccccc}
        1&0 &0 &3 E_{6} & 0&0 &3E_{3} &0 &0  \\
        0 & 1&0 &0 &3 E_{6} & 0&0 &3E_{3} &0 \\
        0 & 0& 1&0 &0 &3 E_{6} & 0&0 &3E_{3} \\
        3E_{3} &0 & 0& 1&0 &0 &3 E_{6} & 0&0 \\
        0& 3E_{3} &0 & 0& 1&0 &0 &3 E_{6} &0 \\
        0 & 0& 3E_{3} &0 & 0& 1&0 &0 &3 E_{6}\\
        3 E_{6} &0 & 0& 3E_{3} &0 & 0& 1&0 &0\\
        0 & 3 E_{6} &0 & 0& 3E_{3} &0 & 0& 1&0\\
        0 &0 & 3 E_{6} &0 & 0& 3E_{3} &0 & 0& 1\\
   \end{array}
   \right),
   $$
   $\hat{p} = \frac{1}{3} \hat{p}_{-1} + \frac{1}{9} \hat{p}_{-2}$,
$$
\hat{p} = \frac{1}{9}
\left(
\begin{array}{ccccccccc}
     4&E_{8} &E_{7} & 3 E_{6}& E_{5}& E_{4}&3 E_{3} &E_{2} & E_{1}\\
     E_{1} & 4 &E_{8} &E_{7} & 3 E_{6}& E_{5}& E_{4}&3 E_{3} &E_{2}\\
     E_{2} &E_{1} & 4 &E_{8} &E_{7} & 3 E_{6}& E_{5}& E_{4}&3 E_{3}\\
     3 E_{3}&E_{2} &E_{1} & 4 &E_{8} &E_{7} & 3 E_{6}& E_{5}& E_{4}\\
     E_{4} &3 E_{3}&E_{2} &E_{1} & 4 &E_{8} &E_{7} & 3 E_{6}& E_{5}\\
     E_{5} &E_{4} &3 E_{3}&E_{2} &E_{1} & 4 &E_{8} &E_{7} & 3 E_{6}\\
     3 E_{6} &E_{5} &E_{4} &3 E_{3}&E_{2} &E_{1} & 4 &E_{8} &E_{7} \\
     E_{7} &3 E_{6} &E_{5} &E_{4} &3 E_{3}&E_{2} &E_{1} & 4 &E_{8} \\
     E_{8} & E_{7} &3 E_{6} &E_{5} &E_{4} &3 E_{3}&E_{2} &E_{1} & 4
\end{array}
\right).
$$
\subsection*{A4. The case $n = 2$ and $N = 3^2 = 9$. Symmetric system.}
$$x=x_0 + 3 x_1 = 
\text{diag}(4;3;2;1;0;-1;-2;-3;-4),$$
$$x_0=\text{diag}(1;0;-1;1;0;-1;1;0;-1),$$
    
$$x_1=\text{diag}(1;1;1;0;0;0;-1;-1;-1).$$
 
Let's denote:
$$
 G_{n} =  \frac{(-1)^n}{2\sqrt{3}\sin\left(\frac{9-n}{9}\right)},
$$
 then:
$$
\hat{p}_{-1} = \frac{1}{\sqrt{3}\I}\left(
\begin{array}{ccccccccc}
     0&G_{8} &G_{7} & 0& G_{5}& G_{4}&0 &G_{2} & G_{1}\\
     G_{1} & 0 &G_{8} &G_{7} & 0& G_{5}& G_{4}&0 &G_{2}\\
     G_{2} &G_{1} & 0 &G_{8} &G_{7} & 0& G_{5}& G_{4}&0\\
     0&G_{2} &G_{1} & 0 &G_{8} &G_{7} & 0& G_{5}& G_{4}\\
     G_{4} &0&G_{2} &G_{1} & 0 &G_{8} &G_{7} & 0& G_{5}\\
     G_{5} &G_{4} &0&G_{2} &G_{1} & 0 &G_{8} &G_{7} & 0\\
     0 &G_{5} &G_{4} &0&G_{2} &G_{1} & 0 &G_{8} &G_{7} \\
     G_{7} &0 &G_{5} &G_{4} &0&G_{2} &G_{1} & 0 &G_{8} \\
     G_{8} & G_{7} &0 &G_{5} &G_{4} &0&G_{2} &G_{1} & 0
\end{array}
\right),
$$
$$
   \hat{p}_{-2} = \frac{1}{\sqrt{3}\I} \left(
   \begin{array}{ccccccccc}
        0&0 &0 &1 & 0&0 &-1  &0 &0  \\
        0 & 0&0 &0 &1 & 0&0 &-1  &0 \\
        0 & 0& 0&0 &0 &1 & 0&0 &-1  \\
        -1 &0 & 0& 0&0 &0 &1 & 0&0 \\
        0& -1  &0 & 0& 0&0 &0 &1 &0 \\
        0 & 0& -1  &0 & 0& 0&0 &0 &1\\
        1 &0 & 0& -1  &0 & 0& 0&0 &0\\
        0 & 1 &0 & 0& -1  &0 & 0& 0&0\\
        0 &0 & 1 &0 & 0& -1  &0 & 0& 0\\
   \end{array}
   \right);
   $$
   $\hat{p} = \frac{1}{3} \hat{p}_{-1} + \frac{1}{9} \hat{p}_{-2}$,
$$
\hat{p} = \frac{1}{9\sqrt{3}\I}\left(
\begin{array}{ccccccccc}
     0&3G_{8} &3G_{7} & 1& 3G_{5}& 3G_{4}&-1&3G_{2} & 3G_{1}\\
     3G_{1} & 0 &3G_{8} &3G_{7} & 1& 3G_{5}& 3G_{4}&-1 &3G_{2}\\
     3G_{2} &3G_{1} & 0 &3G_{8} &3G_{7} & 1& 3G_{5}& 3G_{4}-1\\
     -1&3G_{2} &3G_{1} & 0 &3G_{8} &3G_{7} & 1& 3G_{5}& 3G_{4}\\
     3G_{4} &-1&3G_{2} &3G_{1} & 0 &3G_{8} &3G_{7} & 1& 3G_{5}\\
     3G_{5} &3G_{4} &-1&3G_{2} &3G_{1} & 0 &3G_{8} &3G_{7} & 1\\
     1 &3G_{5} &3G_{4} &-1&3G_{2} &3G_{1} & 0 &3G_{8} &3G_{7} \\
     3G_{7} &1 &3G_{5} &3G_{4} &-1&3G_{2} &3G_{1} & 0 &3G_{8} \\
     3G_{8} & 3G_{7} &1 &3G_{5} &3G_{4} &-1&3G_{2} &3G_{1} & 0
\end{array}
\right).
$$
\subsection*{Appendix B}
Let's consider the plot of the nodes of the lattice for the arbitrary digit $d_1$ (Fig. \ref{systems}). We can see that the $d_1$ is the parameter of the translation over the vector $\mathbf{V} = (1/2, -1)$. Thus we obtain something like a torus winding with step $\Delta x$, and $d_1$ defines the section of the torus.
\begin{figure}[h]
\centering
\begin{tikzpicture}
\draw [->] (0,0) -- (6,0) node [below] {$x$};
\draw [->] (0,0) -- (0,3) node [left] {$d_1$};
\draw [->, thick] (0,0) -- (2,0);
\draw [-, thick] (2,0) -- (4,0) node[below] {4};
\draw [->, thick] (0,2) -- (2,2);
\draw [-, thick] (2,2) -- (4,2);
\draw [->, thick] (0,0) -- (0,1);
\draw [->, thick] (0,1) -- (0,2) node[left] {2};
\draw [->, thick] (4,0) -- (4,1);
\draw [-, thick] (4,1) -- (4,2);
\draw[-, thick, red] (0,2) -- (2,0);
\draw[-, thick, red] (1,2) -- (3,0);
\draw[-, thick, red] (2,2) -- (4,0);
\draw[-, thick, red] (3,2) -- (4,1);
\draw[-, thick, red] (0,1) -- (1,0);
\draw[fill = red] (0,0) circle (1pt);
\draw[fill = red] (4,2) circle (1pt);
\draw[-, blue, dashed, thick] (-0.5,2) -- (4.5,2) node[right] {$d_1 = 0$};
\draw[-, blue, dashed, thick] (-0.5,1) -- (4.5,1) node[right] {$d_1 = -1$};
\draw[-, blue, dashed, thick] (-0.5,1.5) -- (4.5,1.5) node[right] {$d_1 = -1/2$};
\end{tikzpicture}
\caption{The torus winded with the red lines represents the view of the lattice for any possible value of $d_1$ for $\Delta x = 1$, $\Xi = 4$}
\end{figure}
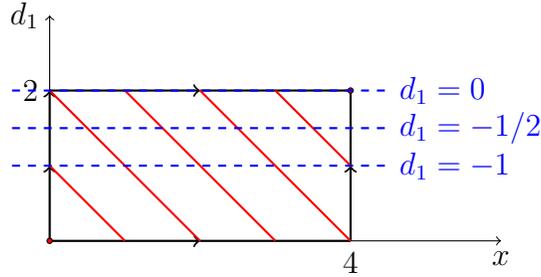
\\
Here we can see, that if we change  $d_1$ to $d_1 + 1$, we obtain the same result.

We obtain a less trivial connection with the windings of torus if we plot over the $x$-axis the momentum and over the $y$-axis the momentum with respect of modulo $1$, in other words -- the fractional part of the momentum. Then we obtain the winding of torus with unite slope. Let $\Pi = 2^{n_-} = 4$, $d_1 = 0$, then we have the following plot:
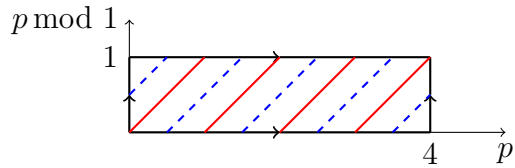
\begin{figure}[h]
\centering
\begin{tikzpicture}
\draw [->] (0,0) -- (5,0) node [below] {$p$};
\draw [->] (0,0) -- (0,1.5) node [left] {$p \, \text{mod} ~ 1$};
\draw [->, thick] (0,0) -- (2,0);
\draw [-, thick] (2,0) -- (4,0) node[below] {4};
\draw [->, thick] (0,1) -- (2,1);
\draw [-, thick] (2,1) -- (4,1);
\draw [->, thick] (0,0) -- (0,0.5);
\draw [-, thick] (0,0.5) -- (0,1) node[left] {1};
\draw [->, thick] (4,0) -- (4,0.5);
\draw [-, thick] (4,0.5) -- (4,1);
\draw[-, thick, red] (0,0) -- (1,1);
\draw[-, thick, red] (1,0) -- (2,1);
\draw[-, thick, red] (2,0) -- (3,1);
\draw[-, thick, red] (3,0) -- (4,1);
\draw[-, thick,  dashed, blue] (0,0.5) -- (0.5,1);
\draw[-, thick,  dashed, blue] (0.5,0) -- (1.5,1);
\draw[-, thick, dashed, blue] (1.5,0) -- (2.5,1);
\draw[-, thick, dashed, blue] (2.5,0) -- (3.5,1);
\draw[-, thick, dashed, blue] (3.5,0) -- (4,0.5);
\end{tikzpicture}
\caption{The plot of fractional part of momentum for $\Pi = 4$, $d_1 = 0$ (red lines), $d_1 = -0.5$ (blue dashed line)}
\end{figure}
\\
Changing the digit $d_1$ we can move this winding without changing it's slope. In particular, the plot is shifted for $-d_1$ over the $p$-axis (or the coordinate axis is shifted for $d_1$).

\end{document}